\documentclass[9pt,onecolumn,twoside]{opticajnl}
\journal{opticajournal} % use for journal or Optica Open submissions

\setboolean{shortarticle}{true}
\usepackage{lineno}
\nolinenumbers
\title{Shaping space-time wavepackets beyond the paraxial limit using a dispersion magnifier}

\author[1]{Dongha Kim}
\author[1]{Cheng Guo}
\author[1]{Peter B. Catrysse}
\author[1,*]{Shanhui Fan}

\affil[1]{E. L. Ginzton Laboratory, Stanford University, 348 Via Pueblo, CA 94305, United States}

\affil[*]{shanhui@stanford.edu}

\begin{abstract}
Space-time wavepackets (STWPs) have received significant attention since they can propagate in free space at arbitrary group velocity without dispersion and diffraction. However, at present, the generation of STWPs has been limited to the paraxial regime. Here we show that conventional optical elements can be used to extend STWPs beyond the paraxial regime. A dispersion magnifier, consisting of two lenses and a beam expander, applies spatiotemporal shaping to paraxial STWPs to create nonparaxial STWPs. The control of the magnification ratio results in versatile engineering capabilities on group velocity, beam diameter, and propagation distance. As an example, we numerically demonstrated long-distance propagation or slow group velocity of the output wavepacket with subwavelength cross-sections.
\end{abstract}

\setboolean{displaycopyright}{false} % Do not include copyright or licensing information in submission.

\begin{document}

\maketitle

Space-time wavepackets (STWPs) are a distinctive class of structured optical pulses possessing shape-preserving properties in space and time with engineered group velocities (\(v_g\)) \cite{Yessenov_AOP22}. These wavepackets can be formed in either free space or dispersive media without the use of nonlinearity. Their formation involves exciting only those plane wave components that have their frequencies (\(\omega\)) and longitudinal wavevectors (\(k_z\)) satisfying a dispersion relation \(\omega=\omega(k_z)\). This dispersion relation is also referred to as space-time coupling. The preparation of space-time coupling has been proposed or demonstrated using various devices, such as multimode optical fibers \cite{Sefanska_ACSPHOTON2023}\cite{Cheng_PRR2021}, spatial light modulators \cite{Kondakci_NP2017}\cite{Yessenov_OE2019}, and photonic crystal slabs \cite{Guo_LSA2021}. In particular, the high programmability of spatial light modulators has enabled the demonstration of various properties of STWPs, including arbitrary group velocity engineering \cite{Kondakci_NCOMMS2019}, anomalous refraction and reflection \cite{Bhaduri_NP2020}, and complete localization of wavepackets in all dimensions \cite{Yessenov_ncomms2022}.

\begin{figure}[b!]
\centering
\includegraphics[width=0.5\linewidth]{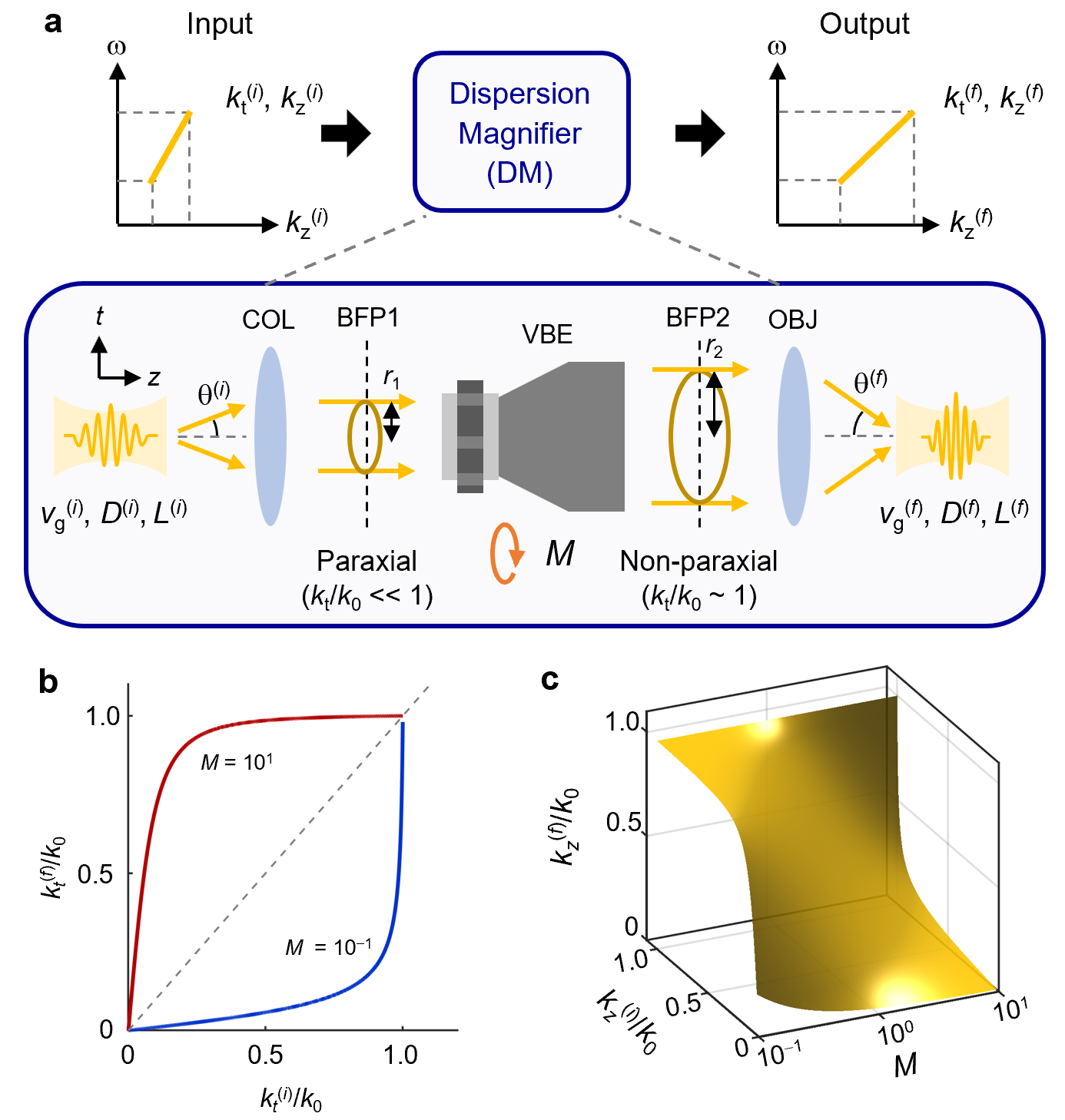}
\caption{(a) Schematic illustration of dispersion magnifier and its operation mechanism. COL (OBJ) is the collimating (imaging) lens, respectively. VBE is the variable beam expander. BFP1 (BFP2) is the backfocal plane of COL (OBJ), respectively. \(M\) is the magnification ratio. (b) Relation between the input and output transverse wavevectors \(k_t^{(i)}/k_0 - k_t^{(f)}/k_0\) for the magnification ratio \(M\) = \(10^{1}\) (red), and \(10^{-1}\) (blue), respectively. The gray dashed line indicates the condition of \(k_t^{(i)}\) = \(k_t^{(f)}\). (c) Relation between the input and output longitudinal wavevector \(k_z^{(i)}/k_0 - k_z^{(f)}/k_0\) depending on \(M = 10^{-1} - 10^1\).}
\label{fig1}
\end{figure}

However, the demonstrations of STWPs has not been achieved in non-paraxial regimes so far \cite{Yessenov_AOP22}. Here, the paraxial regime refers to the scenario where the diffraction angle \(\theta\), i.e., the angle between the wavevector of the plane wave components of the wavepacket and the propagation axis, fulfills \(\theta \approx \sin{\theta}\) with less than 1.5\(\%\) error \cite{Pedrotti}, which corresponds to \(\theta < 10^{\circ}\) or numerical aperture \(NA < 0.17\). By generating STWPs in the non-paraxial regime, long-distance propagation of pulses with wavelength-scale cross-section can be demonstrated, which may enable high-density information processing in free space \cite{Liao_NP2017}\cite{Kang_NCOMMS2019} and transverse spatial resolution enhancement in light sheet fluorescence microscopy \cite{Fahrbach_NCOMMS2012}\cite{Takanezawa_NCOMMS2021}.

In this work, we theoretically propose a spatiotemporal pulse shaping method, which we call a dispersion magnifier, that can generate STWPs in non-paraxial regime and engineer their properties with conventional optical elements (Fig. \ref{fig1}a). For an incident STWP in the paraxial regime, the dispersion magnifier can modify its dispersion relation by a stretching or shrinking transformation along the transverse wavevector (\(k_t\)) axis, allowing access to both the paraxial and non-paraxial regimes. As a result, various properties of STWPs can be manipulated by the dispersion magnifier, including transverse wavevector, group velocity, beam diameter, and propagation distance.

% \section{Fourier-optic dispersion magnification}

The dispersion magnifier consists of a collimating lens, an imaging lens, and a beam expander in between (Figure 1a). For the incident STWP, at each frequency \(\omega\), the plane-wave components have the same longitudinal wavevector \(k_z^{(i)}\). Their transverse wavevectors have the same magnitude \(k_t^{(i)}\). The magnitude of wavevector \(k_0\) is defined as \(k_0 = \sqrt{(k_z^{(i)})^2+(k_t^{(i)})^2}\). After the input STWP passes through the collimating lens, it forms a ring-shaped beam in real space, with each frequency component located at a ring with a radius \(r_1\) of: 

\begin{equation} \label{eqn1}
r_1 = f_C \tan{(\theta^{(i)})} = f_C k_t^{(i)}/k_z^{(i)}.
\end{equation}

\noindent where \(f_C\) is the focal length of the collimating lens and \(\theta^{(i)}\) is the propagating angle of this frequency component in the input STWP. The ring-shaped beam then passes through the variable beam expander. The light at the ring with the radius \(r_1\) is mapped to light at a ring with a radius \(r_2\) as

\begin{equation} \label{eqn2}
r_2 = M r_1.
\end{equation}

\noindent where \(M\) is the magnification ratio of the beam expander. The output from the beam expander then passes through the imaging lens. The magnitude of the transverse wavevector of the output STWP (\(k_t^{(f)}\)) is related to the ring radius \(r_2\) as

\begin{equation} \label{eqn3}
k_t^{(f)} = k_0\sin{(\theta^{(f)})} = \frac{r_2}{\sqrt{r_2^2+f_O^2}}
\end{equation}

\noindent where \(f_O\) is the focal length of the imaging lens and \(\theta^{(f)}\) is the propagating angle of this frequency component in the output STWP. As a result, the relations between the input and output wavevector components can be derived by combining Eqs. \ref{eqn1}, \ref{eqn2} and \ref{eqn3} as:

\begin{align} 
k_t^{(f)}/k_0 &= \frac{M (k_t^{(i)}/k_0)}{\sqrt{1+(M^2-1)(k_t^{(i)}/k_0)^2}} \label{eqn4} \\
k_z^{(f)}/k_0 &= \frac{k_z^{(i)}/k_0}{\sqrt{M^2+(1-M^2)(k_z^{(i)}/k_0)^2}} \label{eqn5}
\end{align}

\noindent where we assumed \(f_C = f_O\) for simplicity. 

The transformation in Eq. \ref{eqn4} allows one to transform a STWP between the paraxial and non-paraxial regimes. As an illustration, in Fig. \ref{fig1}b, we plot the relation of  Eq. \ref{eqn4} for two different magnification ratios. The paraxial regime corresponds to \(k_t/k_0 < 0.17\). Here and throughout the paper, we consider the magnification ratio \(M\) to be in the range of \(0.1< M <10\) for a dispersion magnifier, a range that is accessible in bulk lens systems \cite{salehteich}. For \(M\) = 10,  an incident STWP with \(k_t^{(i)}/k_0\) in the range of [0.017, 0.17], which is in the paraxial regime, will be transformed into a wavepacket outside the paraxial regime. Conversely, for \(M\) = 0.1, an incident STWP with \(k_t^{(i)}/k_0\) in the range of [0.17, 0.865], which is outside the paraxial regime, will be transformed into a wavepacket within the paraxial regime. In Fig. \ref{fig1}c, we plot the relation of Eq. \ref{eqn5} for different magnification ratios. When \(M \ne 1\), the longitudinal wavevectors transform, resulting in the change of space-time coupling and hence the group velocity of the STWP. Related to our work, the manipulation of space-time coupling using conventional optical elements has been considered in \cite{Porras_PRA2018}\cite{Li_SR2022}. Ref. \cite{Porras_PRA2018}\cite{Li_SR2022}, however, are restricted to the paraxial regime. The implications of these elements for non-paraxial regime has not been previously analyzed.  

% \section{Magnifying the dispersion of STWPs}

\begin{figure}[t!]
\centering
\includegraphics[width=0.5\linewidth]{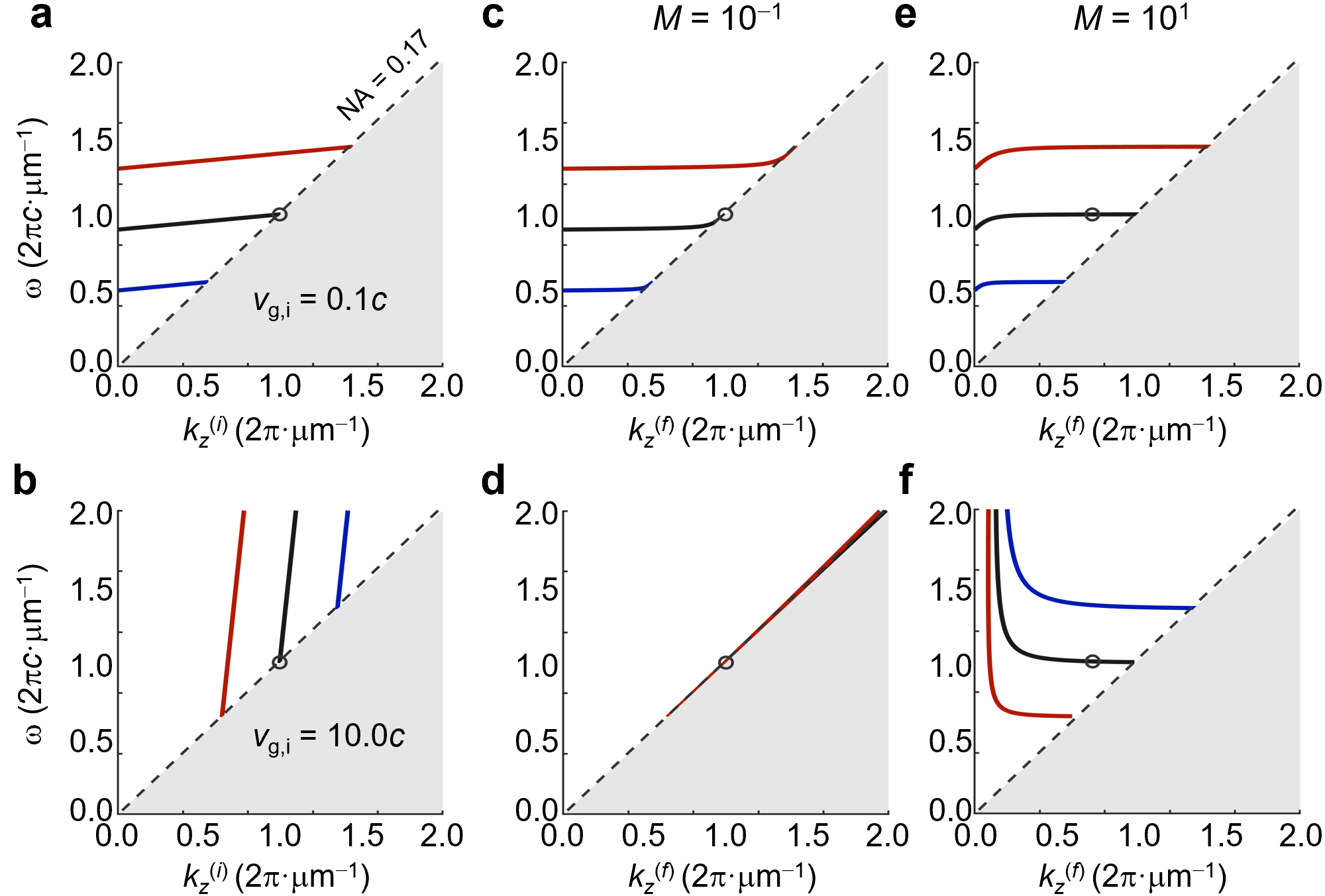}
\caption{(a) Dispersion of input STWPs (\(\omega = v_g^{(i)} k_z + \omega_0\)) with subluminal group velocity \(v_g^{(i)} = 0.1\) and \(\omega_0\) = \(1.5\) (red), \(1.0\) (black), \(0.5\) (blue). (b) Dispersion of input STWPs with superluminal group velocity \(v_g^{(i)} = 10.0\) and \(\omega_0\) = \(-2\) (red), \(-6\) (black), \(-10\) (blue). (c,d) Dispersion of output STWPs at magnification \(M\) = \(10^{-1}\) for (c) subluminal and (d) superluminal \(v_g^{(i)}\). (e,f) Dispersion of output STWPs at magnification \(M\) = \(10^{1}\) for (e) subluminal and (f) superluminal \(v_g^{(i)}\). The black dashed line indicates the paraxial limit (\(NA = 0.17\)). The black circles indicate the wavelength of 800 nm. The white (gray-shaded) region corresponds to the regions above (below) the lightline. The units of \(v_g\) and \(\omega_0\) are \(c\) and \(2\pi c \cdot \mu m^{-1}\), respectively.}
\label{fig2}
\end{figure}

In Fig. \ref{fig2}, we show how the space-time couplings of the STWPs are transformed by the dispersion magnifier. The space-time coupling of any of these STWPs lies above the light line of \(\omega = c k_z\). With the scale of Fig. \ref{fig2}, the paraxial regime corresponds to a very narrow slice in the immediate vicinity of the light line. Thus, most of the plots here concern non-paraxial regime. We consider the input wavepackets with a space-time coupling of \(\omega = v_g^{(i)} k_z+\omega_0\). \(\omega_0 = \omega_c (1-\frac{v_g}{c}\cos(\theta_c))\), where \(\omega_c\) is the central frequency of the spectrum and \(\theta_c\) is the propagating angle of the planewave component at \(\omega = \omega_c\). The \(v_g^{(i)}\) is chosen to be subluminal at \(0.1c\) and superluminal at \(10.0c\) in Figs. \ref{fig2}a and \ref{fig2}b, respectively. The space-time coupling for the corresponding output wavepackets, for two magnification ratios of \(M = 10^{-1}\) and \(10^1\), are shown in Figs. \ref{fig2}c-f. In general, the space-time coupling for the output STWP is no longer linear. For \(M = 10^{-1}\), a significant part of the space-time coupling is pushed towards the lightline (Fig. \ref{fig2}c-d). For \(M = 10^1\), the space-time coupling, in general, is pushed away from the lightline. (Fig. \ref{fig2}e-f). With the superluminal input wavepacket (Fig. \ref{fig2}b), at \(M = 10^1\), the output may exhibit a negative group velocity (blue curve, Fig. \ref{fig2}f). The results here indicate that a rich set of space-time coupling behavior can be generated, especially in the non-paraxial regime.

\begin{figure}[t!]
\centering
\includegraphics[width=0.5\linewidth]{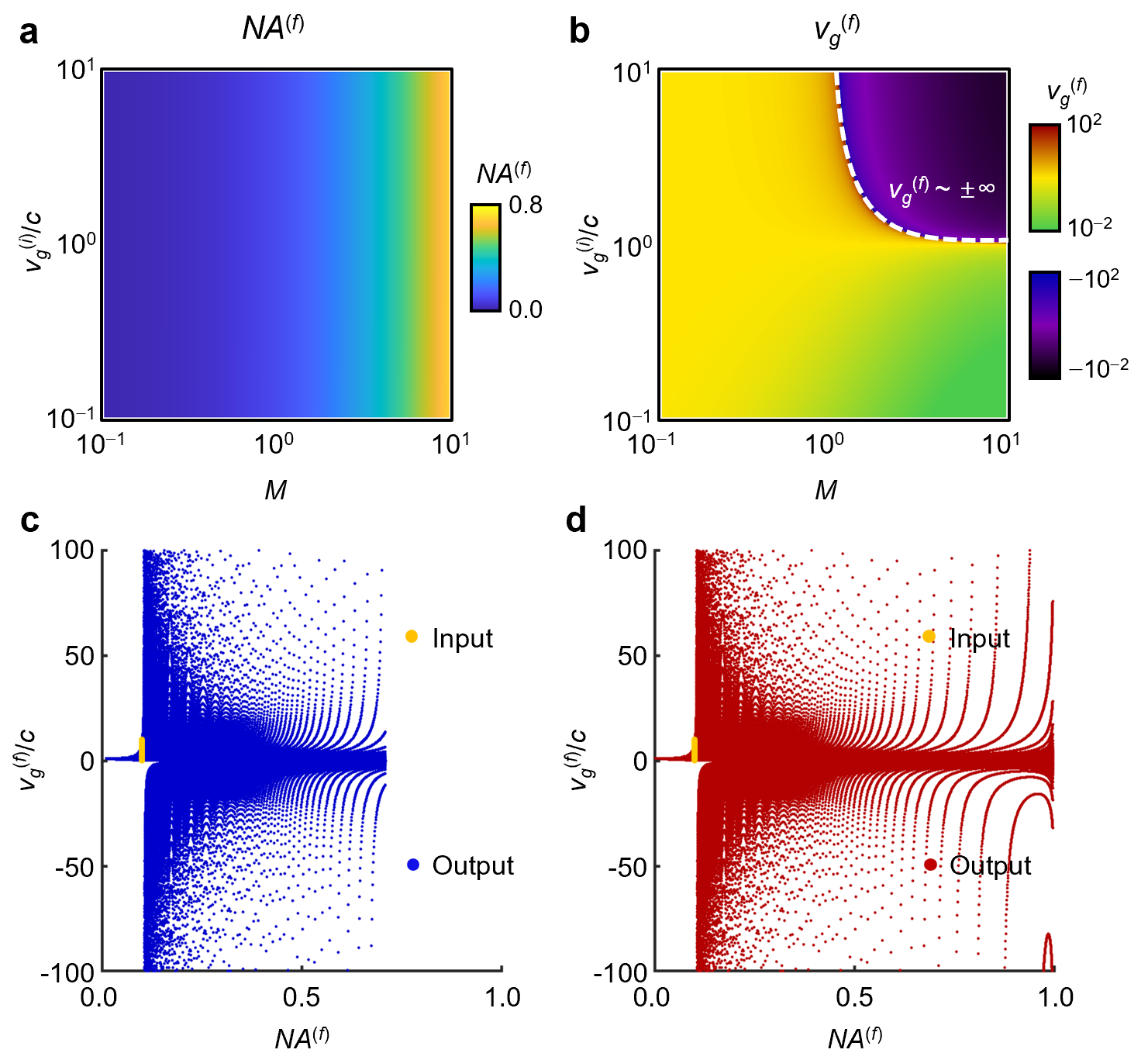}
\caption{(a) Calculated maximum numerical aperture (\(NA^{(f)}\)) and (b) group velocity \(v_g^{(f)}\) of output STWPs for incident paraxial STWP (\(NA^{(i)} = 0.1\)) with group velocities of \(v_g = 10^{-1}c - 10^{1}c\). (c,d) Achievable combinations of \(v_g^{(f)}\) and \(NA^{(f)}\) in (c) a single dispersion magnifier and (d) two cascaded dispersion magnifiers.}
\label{fig3}
\end{figure}

We investigate the engineering of the numerical aperture and group velocity of STWPs in dispersion magnifier (Fig. \ref{fig3}). We assume the same kind of input wavepackets as described in Fig. \ref{fig2}, with a center wavelength of \(\lambda\) = 0.8 \(\mu\)m, the group velocities in the range of \(v_g = 0.1c-10c\), and a maximum numerical aperture of \(NA = 0.1\).  We further assume that these wavepackets are sufficiently narrow-banded, such that the space-time coupling of the output wavepacket can be approximated as linear. This narrow-band assumption is, in practice, applicable to most existing experiments on STWP \cite{Sefanska_ACSPHOTON2023}\cite{Yessenov_ncomms2022}. For \(v_g\) = \(0.1c\) and \(10c\), the input wavepackets are represented by the black circles in Figure \ref{fig2}a and \ref{fig2}b, respectively. The calculated maximum numerical aperture \(NA^{(f)} = k_t^{(f)}/k_0\) of output STWPs varies from 0.01 to 0.71, covering both paraxial and nonparaxial regimes depending on \(M\) (Fig. \ref{fig3}a). The NA engineering is independent of the incident group velocities. The calculated group velocity \(v_g^{(f)}/c\) of the output STWPs can be dramatically changed from subluminal to superluminal (\(3.1\times10^{-3}c-5.9\times10^{3}c\)), even including the negative group velocities from \(-1.5\times10^{4}c\) to \(-3.2\times10^{-2}c\) (Fig. \ref{fig3}b). 

We find all possible combinations of \(NA^{(f)}-v_g^{(f)}\) that can be achieved in the dispersion magnifier.
In Figs. \ref{fig3}c and \ref{fig3}d, the input wavepackets are represented by yellow dots. Fig. \ref{fig3}c considers the use of a single dispersion magnifier with \(0.1 < M < 10\).  For \(M < 1\), the numerical aperture of the output wavepacket \(NA^{(f)} < NA^{(i)}\). The output wavepacket remains in the paraxial regime, and the group velocity of the output wavepacket resides in a narrower range as compared with that of the input wavepackets. For \(M > 1\), \(NA^{(f)} > NA^{(i)}\). The output wavepacket can be in the non-paraxial regime, and the group velocity of the output wavepacket takes a much wider range spanning from \(-100c\) to \(100c\). In Fig. \ref{fig3}c, with the magnification ratio \(M\) restricted to less than 10, the maximum numerical aperture that can be achieved for the output wavepacket is restricted to \(NA^{(f)} < 0.71\). To achieve a higher numerical aperture, one can cascade two dispersion magnifers, resulting in the total magnification ratio of  \(0.01 < M < 100\). The results for two cascaded dispersion magnifiers are shown in Figure \ref{fig3}d with the same input wavepackets. We see that the numerical aperture for the output wavepacket can reach unity in this case. Such capability for reaching high numerical aperture is important for creating wavelength-scale STWPs, as we will discuss explicitly below. 

% \section{Engineering beam diameter and propagation distance}

In the discussion above, we assume that the spectrum of a light bullet is along a line with zero thickness in the \(\omega-k_z\) space, as shown in Figure \ref{fig2}. Under such an assumption, a light bullet can propagate indefinitely without any distortion but such a light bullet has infinite energy. In practice, the spectrum of a light bullet has a non-zero spread along the \(k_z\) direction for each \(\omega\). Such a non-zero spread is necessary for the light bullet to have finite energy. For such a practical light bullet, the propagation distance is no longer infinite. Assuming axial symmetry around the propagation axis \(z\), we consider an incident STWP at \(z = 0\) with its complex amplitudes of individual plane-wave components \(\tilde A\) described as,

\begin{equation}
        \tilde A(k_x, k_y, z=0, \omega) = \exp{\left(-\frac{(k_z - k_{z,c}(\omega))^2}{2 \delta k_z^2 (\omega)}\right)} \exp{\left(-\frac{(\omega - \omega_{c})^2}{2 \Delta \omega^2}\right)} \nonumber
\end{equation}

\noindent where \(k_z=\sqrt{\omega^2/c^2 -k_x^2-k_y^2}\), \(k_{z,c}(\omega)\) is the average space-time correlation for the wavepacket. \(\omega_c\) is the central frequency of the spectrum. \(\Delta \omega \ll \omega_c\) is the frequency bandwidth of the spectrum, and is related to the wavelength bandwidth \(\Delta \lambda\) of the spectrum. \(\delta k_z(\omega)\) is the linewidth of the longitudinal wavevector \(k_z\), which is related to the frequency linewidth \(\delta \omega\) of the space-time coupling as:

\begin{equation}
    \delta k_z(\omega) = \frac{\delta\omega}{c \cos{(\theta_c)}} \left(1- \frac{c}{v_g} \cos{(\theta_c)} \right). \label{eqn6}
\end{equation}

\noindent as described in Section I of the Supplementary Information.  Here we assume that \(\delta \omega \ll \omega\), which implies that  \(\delta k_z(\omega) \ll k_z(\omega)\). The pulse \(A(x,y,z,t)\) of STWPs after propagation over a distance \(z\) is,

\begin{align}
        A(x,y,z,t) =  \iiint  \,d\omega\,dk_x\,dk_y\ \tilde A(k_x,k_y,0,\omega) 
        e^{\left(i[k_x x + k_y y + k_z z -\omega t]\right)} \label{eqn7} 
\end{align}

\noindent We define the beam diameter \(D\) as the full width at half maximum of the intensity distribution along the transverse spatial axis at \(z = 0\) and \(t = 0\). The propagation distance \(L\) is defined as the distance at which the maximum intensity attenuated to half of its value at \(z = 0\). At any given z, the maximum intensity of the pulse is always located at \(x = y = 0\), and reached at the time \(t = z/v_g\). 

\begin{figure}[t!]
\centering
\includegraphics[width=0.5\linewidth]{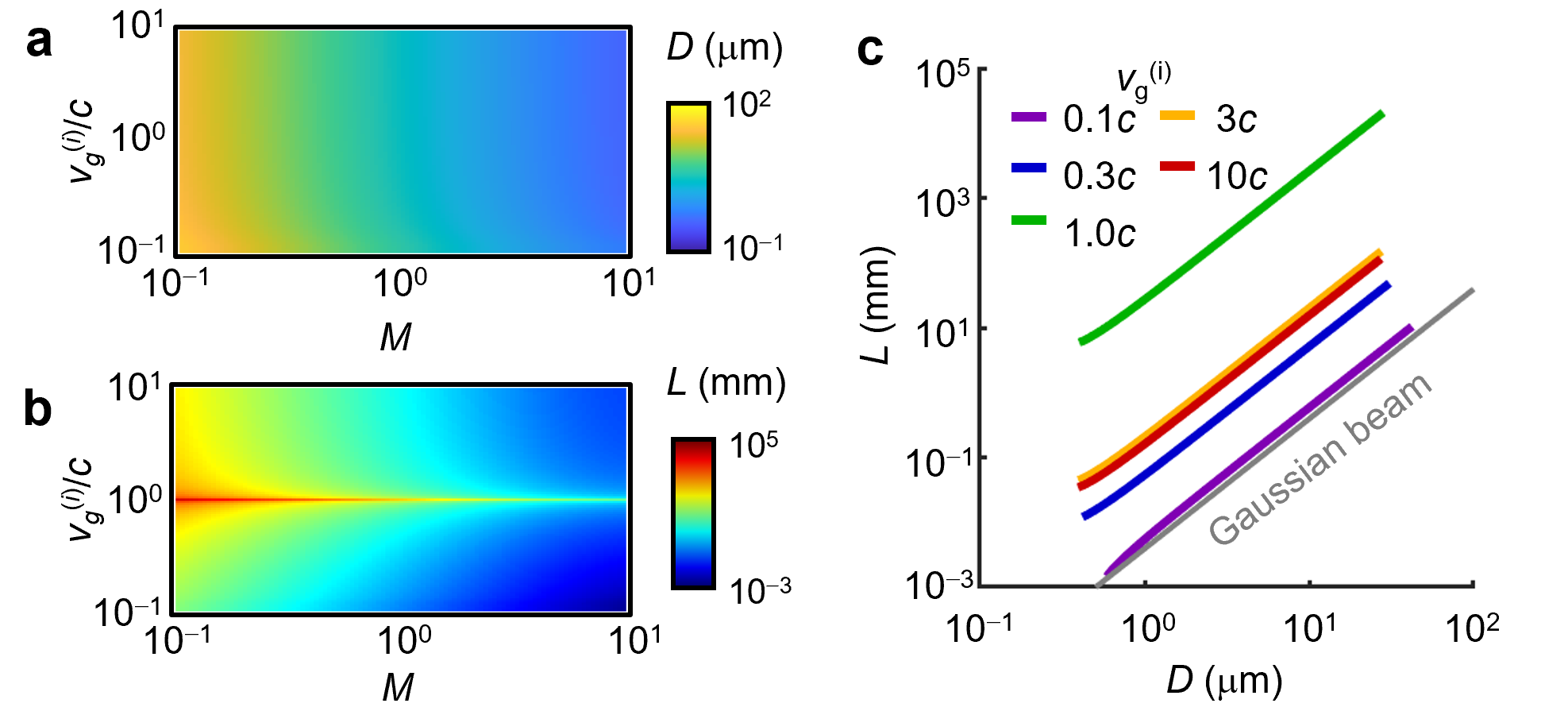}
\caption{(a) Calculated beam diameter (\(D\)) and (b) propagation distance (\(L\)) of output STWPs. The upper insets illustrate the definition of \(D\) and \(L\). (c) Combinations of \(D\) and \(L\) of output STWPs depending on the group velocity of input STWPs (\(v_g^{(i)}\)). The gray line presents the \(D-L\) relation for a Gaussian beam with 800-nm-wavelength.}
\label{fig4}
\end{figure}

% (a) Calculated beam diameter (\(D\)) and (b) propagation length (\(L\)) of output STWPs. The upper insets illustrate the definition of \(D\) and \(L\). (c) Calculated average linewidths of the longitudinal wavevector (\(\Delta k_z\)) depending on the initial group velocity (\(v_g^{(i)}\)). The orange line is the \(\Delta k_z\) before the dispersion magnifier. The blue and yellow lines are the \(\Delta k_z\) after the dispersion magnifier with \(M = 10^{-1}\) and \(M = 10^1\), respectively. (d) Achievable combinations of \(v_g^{(f)}\) and \(NA^{(f)}\) in a single dispersion magnifier. The gray line indicates the \(D-L\) relation for a gaussian beam with 800-nm-wavelength.

We calculate the beam diameter \(D\) and the propagation distance \(L\) of output STWPs (Figs. \ref{fig4}a,b). We use the same setup as in Figure \ref{fig3} and consider the output resulting from an input STWP passing through the dispersion magnifier. We assume the finite bandwidth of frequencies and wavevectors, which are defined by a wavelength bandwidth \(\Delta \lambda\) = 0.5 nm and a frequency linewidth \(\delta \omega\) = \(10^{-4} \omega\). Figure \ref{fig4}a present the calculated beam diameter \(D\) for different \(v_g^{(i)}\) and \(M\). The beam diameter can vary from the sub-millimeter-scale (\(10^{2} \mu\)m) to the diffraction-limit of center wavelength of the wavepacket (\(10^{-1} \mu\)m). The distribution of beam diameter is mainly determined by the magnification ratio \(M\), which controls the numerical aperture \(NA^{(f)}\) of the output pulse. Figure \ref{fig4}b presents the calculated propagation distance \(L\) for different \(v_g^{(i)}\) and \(M\). Unlike the beam diameter, \(L\) depends significantly on both \(M\) and \(v_g^{(i)}\), varying from \(10^{-3}\) to \(10^5\) mm. 

The distribution of \(L\) in Figure \ref{fig4}b does not match with the formula \(L = c/\delta\omega/|1-c/v_g|\) as derived in \cite{Yessenov_OE2019}\cite{Guo_LSA2021} since we are mostly outside the paraxial regime. Instead, the propagation distance \(L\) can be expressed as  (See Supplementary Information Section II for derivation),

\begin{equation}
    L \approx \frac{1}{\sqrt{2}dk_z(\omega_c)} = \frac{1}{\sqrt{2}}\frac{c}{\delta\omega} \frac{\cos{(\theta_c)}}{|1-c/v_g \cos{(\theta_c)}|} \label{eqn8}
\end{equation}

\noindent Eq. \ref{eqn8} agrees well with numerical results, as shown in the Supplementary Information Section II.

We analyze the relation between the beam diameter \(D\) and the propagation distance \(L\) for the output STWPs. We consider the same set of input STWP as used in Figure \ref{fig4}a, with \(v_g^{(i)}\) varying from \(0.1c\) to \(10c\). For each input STWP, we vary the magnification ratio \(M\) to vary the output beam diameter \(D\). The resulting propagation distance for the output wavepacket is shown in Figure \ref{fig4}c.  For every \(v_g^{(i)}\), the propagation distance \(L\) increases with the beam diameter \(D\). For the same beam diameter, the propagation distance \(L\) for all output wavepacket is significantly larger than that of a Gaussian wavepacket. For the same beam diameter, Eq. \ref{eqn8} indicates that the propagation distance should depend strongly on the final group velocity \(v_g^{(f)}\). Here we see a strong dependency on the input group velocity \(v_g^{(i)}\) since these two group velocities are directly related as shown in Figure \ref{fig3}b. 

\begin{figure}
\centering
\includegraphics[width=0.5\linewidth]{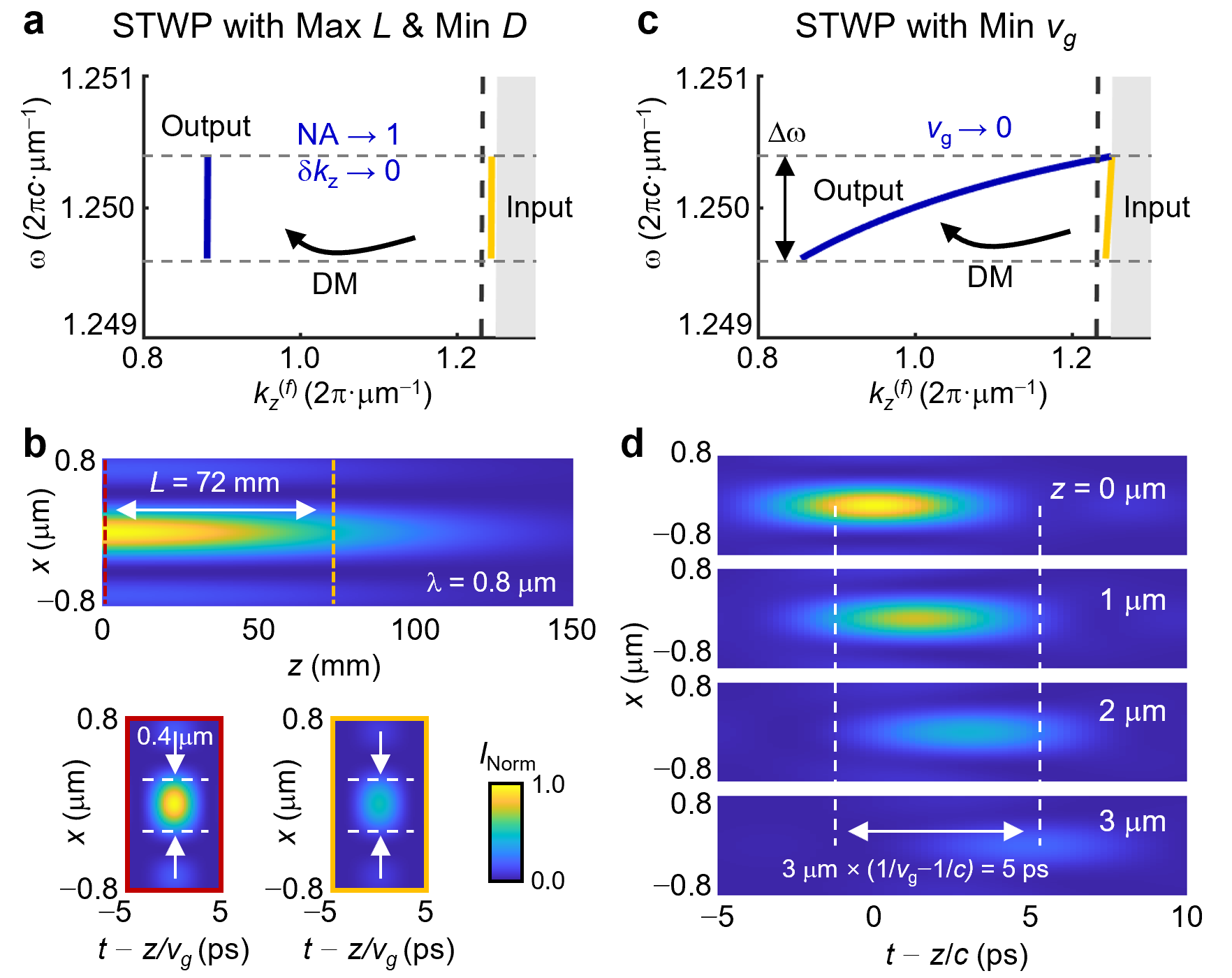}
\caption{(a) Dispersion of input (yellow line) and output STWP (blue line) for long-distance propagation of optical pulses with sub-wavelength-scale cross-section. (b) Numerical demonstration of the output STWP in Figure \ref{fig5}a. (upper) Calculated transverse intensity profile along \(z\)-axis at the temporal locations of maximum peak intensity \(|A(x,0,z,z/v_g)|^2\). The red (orange) dashed line indicates the beam waist (half attenuated distance), respectively. (lower) Calculated beam profile \(|A(x,0,z,t-z/v_g)|^2\) at beam waist (red box) and half attenuated distance (orange box), respectively. (c) Dispersion of input (yellow line) and output STWP (blue line) for optical pulses with slow group velocity. (d) Numerical demonstration of output STWP in Figure \ref{fig5}c. Calculated beam profiles \(|A(x,0,z,t - z/c)|^2\) at \(z\) = 0, 1, 2, 3 \(\mu\)m, respectively. In Figure \ref{fig5}a and c, the black dashed line indicates the paraxial limit (\(NA = 0.17\)). The white (gray-shaded) region corresponds to the regions above (below) the lightline. The units of \(v_g\) and \(\omega_0\) are \(c\) and \(2\pi c \cdot \mu m^{-1}\), respectively.}
\label{fig5}
\end{figure}

% \section{Application to long-distance transport and slow light generation}

Finally, we show the capability of the dispersion magnifier to create STWP with unusual characteristics. In Figure \ref{fig5}a, b, we numerically demonstrate long-distance propagation of optical pulses with subwavelength cross-sections. For this purpose, we take an input wavepacket used in Figure \ref{fig4} in the paraxial limit with a \(NA\) of 0.1, with \(v_g^{(i)} = 0.9954c\), which has a space-time coupling very close to the lightline as indicated by the yellow curve in Figure \ref{fig5}a. We pass the input wavepacket through a dispersion magnifer with \(M = 10\).  The output space-time coupling is significantly away from the lightline, has a much larger numerical aperture. (Blue curve, Figure \ref{fig5}a). The output wavepacket has a sub-wavelength cross-section, yet the sub-wavelength confinement persists even after propagation of approximately 100 mm. 

In Figure \ref{fig5}c and d, we numerically demonstrate the creation of optical pulses with slow group velocity. In this case, we use an input wavepacket with a \(NA\) of 0.1, and with a \(v_g^{(i)} = 0.1c\). Its space-time coupling is in the vicinity of the light line but has a slope that is different from the light line as shown as the yellow curve in Figure \ref{fig5}c. We pass such an input wavepacket through a dispersion magnifier with \(M = 10\). The space-time coupling of the output wavepacket is shown as the blue curve in Figure \ref{fig5}c. It features a group velocity \(v_g^{(f)} = 2.970\times 10^{-3}c\)  (In general, for a given \(M\) there is a complex and non-monotomic dependency of \(v_g^{(f)}\) on \(v_g^{(i)}\) as noted in Figure \ref{fig3}b). The propagation of the output wavepacket is shown in Figure \ref{fig5}d. In this plot, with the choice of the horizontal coordinate, a pulse moving at the speed of light will have a constant center location. The shift in the center location as a function of \(z\) indicates a slow light effect. From the plot, one can infer a group velocity of \(v_g = 1.997\times 10^{-3}c\), in agreement with the space-time coupling shown in Figure \ref{fig5}c.

% \section{Conclusion}
In conclusion, we theoretically investigate the use of the dispersion magnifier for the spatiotemporal shaping of space-time wavepackets. For an input wavepacket in the paraxial regime, the output wavepackets can be tuned to be in either paraxial or non-paraxial regimes, and exhibit a large range of group velocity, beam diameter, and propagation length. Furthermore, we numerically demonstrate long-distance propagation of output wavepacket with subwavelength cross-sections, as well as the creation of the output wavepacket with ultra-slow group velocity. Our study provides a general approach for spatiotemporal pulse shaping of space-time wavepacket, with implications for classical/quantum information processing \cite{Walsh2022_SR}\cite{McLaren2012_OE} and in controlling nonlinear light-matter interactions \cite{Soljacic2002_JOSAB}\cite{Corcoran2009_NP}\cite{Ornelas2024_NP}.

\begin{backmatter}
\bmsection{Funding} The work is supported by a grant from the U. S. Army Research Office (Grant No. W911NF-24-2-0170). D. K. acknowledges the support from the National Research Foundation of Korea (NRF, RS-2023-00240304).

\bmsection{Disclosures} The authors declare no conflicts of interest.

\bmsection{Data Availability Statement} Data underlying the results presented in this paper are not publicly available at this time but may be obtained from the authors upon reasonable request.

\end{backmatter}

% Bibliography
\bibliography{Optica-journal-template}

\newpage
\clearpage
\setcounter{equation}{0}
\setcounter{figure}{0}
\setcounter{table}{0}
\setcounter{page}{1}

\maketitle 
\makeatletter
\renewcommand{\theequation}{S\arabic{equation}}
\renewcommand{\thefigure}{S\arabic{figure}}
\renewcommand{\thetable}{S\arabic{table}}

\begin{center}
\vspace{10pt}
\textbf{\large Supplemental Material for \\``Shaping space-time wavepackets beyond the paraxial limit using a dispersion magnifier''}
\end{center} 
\begin{center} 
{Dongha Kim, Cheng Guo, Peter B. Catrysse and Shanhui Fan$^{\ \textcolor{red}{\dagger}}$}\\
\emph{E. L. Ginzton Laboratory, Stanford University, 348 Via Pueblo, CA 94305, United States}
\vspace{5pt}
\end{center}

% \tableofcontents
\section{Linewidth of the longitudinal wavevector of STWPs}
\label{S1}
\begin{figure}[b]
\centering
\includegraphics[width=0.25\linewidth]{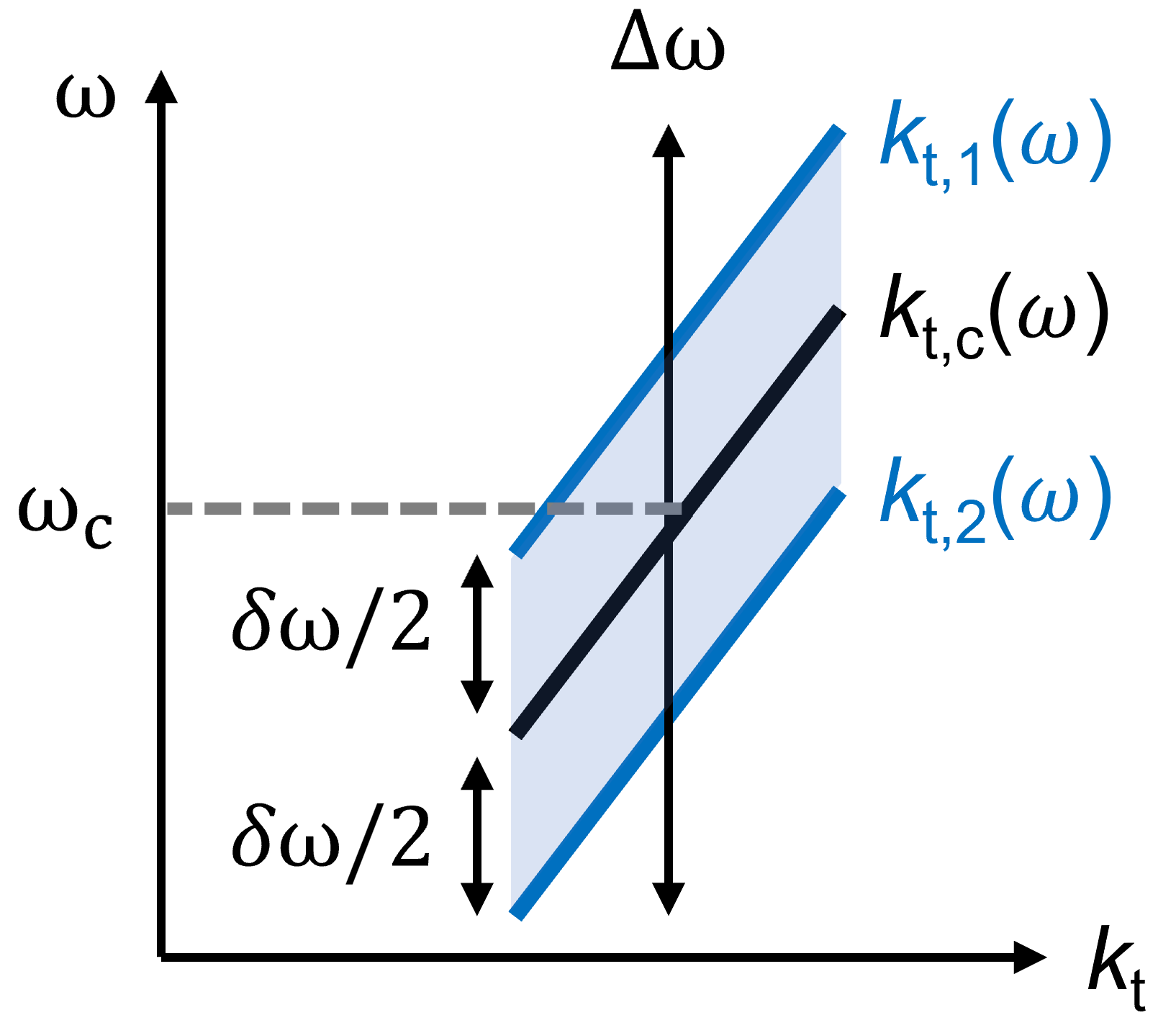}
\caption{Schematic illustration of space-time coupling with finite frequency linewidth \(\delta \omega\) in \(\omega-k_t\) space. The black curve (\(k_{t,c}(\omega))\) indicates the average space-time coupling. The blue curves (\(k_{t,1}(\omega)\) and \(k_{t,2}(\omega)\)) indicate the space-time coupling with the frequency shift of \(\pm \delta \omega /2\), respectively. \(\omega_c\) is the central frequency of the spectrum. \(\Delta \omega\) is the frequency bandwidth of the pulse.}
\label{figS1}
\end{figure}

We derive the linewidth of the longitudinal wavevector of STWPs. First, we define the frequency linewidth \(\delta \omega\) of space-time coupling in \(\omega-k_t\) space, which is described in Figure \ref{figS1}a. The average space-time coupling \(k_{t,c}(\omega)\) is plotted as a black line. The \(k_{t,\pm}(\omega)\) are the space-time coupling with the frequency shift of \(\delta \omega /2\) from the average space-time coupling, which are plotted as blue lines. The \(k_{t,c}(\omega)\) and \(k_{t,\pm}(\omega)\) can be expressed as,

\begin{align}
    k_{t,c}^2(\omega) &= \left( \frac{\omega}{c} \right)^2 -\left( \frac{\omega-\omega_0}{v_g} \right)^2 \nonumber\\
    k_{t,\pm}^2(\omega) &= \left( \frac{\omega \pm \delta \omega}{c} \right)^2 -\left( \frac{\omega-\omega_0 \pm \delta \omega}{v_g} \right)^2 \nonumber
\end{align}

\noindent where \(\omega_0 = \omega_c (1-\frac{v_g}{c}\cos{(\theta_c)})\) and \(\theta_c\) is the propagating angle of planewave component with frequency \(\omega_c\). The space-time coupling can be described in \(\omega-k_z\) space as shown in Figure \ref{figS1}b by coordinate transformation as, 

\begin{align}
    k_{z,c}^2(\omega) &= \left( \frac{\omega-\omega_0}{v_g}\right)^2  \label{eqns1} \\
    k_{z,\pm}^2(\omega) &= \left( \frac{\omega}{c} \right)^2 - \left( \frac{\omega \pm \delta \omega}{c} \right)^2 +\left( \frac{\omega-\omega_0 \pm \delta \omega}{v_g} \right)^2. \label{eqns2}
\end{align}

\noindent At \(\omega = \omega_c\), Eq. \ref{eqns1} becomes

\begin{align}
    k_{z,c}(\omega_c) &= \frac{\omega_c}{c} \cos{(\theta_c)}  \label{eqns3}
\end{align}

\noindent and to the first order of \(\delta \omega\),

\begin{eqnarray}
    k_{z,\pm}(\omega_c) &\approx \frac{\omega_c}{c}\cos{(\theta_c)} \left( 1 \pm \frac{\delta\omega}{\omega_c\cos{(\theta_c)}^2} \left( \frac{c}{v_g} \cos{(\theta_c)} - 1 \right) \right)\label{eqns4}
\end{eqnarray}

\noindent By combining Eq. \ref{eqns3} and \ref{eqns4}, the linewidth of the longitudinal wavevector at central frequency \(\omega_c\) can be obtained as,
\begin{equation}
    \delta k_{z}(\omega_c) = 2|k_{z,\pm}(\omega_c)-k_{z,c}(\omega_c)| \approx \frac{\delta\omega}{c\cos{(\theta_c)}} \left| 1 - \frac{c}{v_g} \cos{(\theta_c)}\right|. \label{eqns5}
\end{equation}

\newpage
\clearpage

\section{Derivation of propagation distance of STWPs}
\label{S2}

We provide a derivation of the propagation distance of STWPS without making the paraxial approximation. As the maximum intensity of the pulse is always located at \(x = y = 0\) and reached at the time \(t = z/v_g\), the longitudinal distribution of maximum amplitude \(A(0,0,z,z/v_g)\) can be expressed as, 

\begin{equation}
    A(0,0,z,z/v_g) = \int_0^{\infty} d\omega \int_{-\infty}^{\infty} \int_{-\infty}^{\infty} dk_x dk_y \exp{\left(-\frac{(\omega - \omega_{c})^2}{2\Delta \omega^2}\right)}  \exp{\left(-\frac{(k_z - k_{z,c}(\omega))^2}{2 \delta k_z^2 (\omega)}\right)} \exp{\left(i\left(zk_z-\frac{z}{v_g}\omega\right)\right)} \label{eqns6}
\end{equation}

\noindent The coordinate of wavevector space can be transformed from \((k_x,k_y)\) to \((k_z,\phi)\) as, \((k_x,k_y) = (\sqrt{(\omega/c)^2 - k_z^2}\cos{\phi},\sqrt{(\omega/c)^2 - k_z^2}\cos{\phi})\) and \(dk_x dk_y = -k_z dk_z d\phi\). Eq. \ref{eqns6} becomes,

\begin{equation}
    A(0,0,z,z/v_g) = \int_0^{\infty} d\omega \exp{\left(-\frac{(\omega - \omega_{c})^2}{2\Delta \omega^2}-i\frac{z}{v_g}\omega\right)} \left[ \int_0^{\infty} \int_0^{2\pi} k_z dk_z d\phi \exp{\left(-\frac{(k_z - k_{z,c}(\omega))^2}{2 \delta k_z^2 (\omega)}+izk_z\right)}\right]. \label{eqns7}
\end{equation}

\noindent The integration in Eq. \ref{eqns7} can be carried out to yield,

\begin{equation}
    \int_0^{\infty} \int_0^{2\pi} k_z dk_z d\phi \exp{\left(-\frac{(k_z - k_{z,c}(\omega))^2}{2\delta k_z^2 (\omega)}+izk_z\right)} = 2\pi  \exp{(ik_{z,c}(\omega)z-\frac{1}{2}\delta k_z^2(\omega)z^2)} \left[2\sqrt{\pi}\delta k_z^2(\omega)C(\omega)\right] \label{eqns8}
\end{equation}

\noindent where \(C(\omega) = (k_{z,c}(\omega)+iz\delta k_z^2(\omega))/\sqrt{2}\delta k_z(\omega)\). By combining Eq. \ref{eqns7}-\ref{eqns8} and \(\omega-\omega_0 = v_g k_z\), Eq. \ref{eqns7} becomes,

\begin{equation}
    A(0,0,z,z/v_g) = \int_0^{\infty} d\omega \exp{\left(-\frac{(\omega - \omega_{c})^2}{2 \Delta \omega^2}-i\frac{z}{v_g}\omega_0\right)} \left( 4\pi^{3/2} \exp{\left(-\frac{1}{2}\delta k_z^2(\omega)z^2\right)} \delta k_z^2(\omega) C(\omega) \right) \label{eqns9}
\end{equation}

\noindent Based on narrow band assumption, the dispersion of linewidths of the longitudinal wavevector can be approximated as constant, as \(\delta k_z(\omega) \approx \delta k_z(\omega_c)\). Then, Eq. \ref{eqns9} becomes,

\begin{equation}
    A(0,0,z,z/v_g) = 4\sqrt{2}\pi^2 \delta k_z^2(\omega_c) \left(\frac{\omega_c}{c}\cos{(\theta_c)}+iz\delta k_z^2(\omega_c)\right) \exp{\left(-\frac{1}{2}\delta k_z^2(\omega_c)z^2\right)} \label{eqns10}
\end{equation}

\noindent and the intensity distribution \(|A(0,0,z,z/v_g)|^2\) is,

\begin{equation}
    |A(0,0,z,z/v_g)|^2 = 32\pi^4\delta k_z^4(\omega_c) \left(\frac{\omega_c^2}{c^2} \cos^2{(\theta_c)} + \delta k_z^4(\omega_c) z^2\right) \exp{(-\delta k_z^2(\omega_c) z^2)}. \label{eqns11}
\end{equation}

\noindent As \(\delta k_z(\omega_c) \ll (\omega_c/c)\cos{(\theta_c)}\), Eq. \ref{eqns11} can be approximated as,

\begin{equation}
    |A(0,0,z,z/v_g)|^2 \approx 32\pi^4\delta k_z^4(\omega_c) \frac{\omega_c^2}{c^2} \cos^2{(\theta_c)}\exp{(-\delta k_z^2(\omega_c) z^2)} \label{eqns12}
\end{equation}

\noindent Based on Eq. \ref{eqns12}, the propagation distance of STWPs is defined as the distance \(L\) where the intensity attenuates into half compared to \(z = 0\), which is expressed as,

\begin{equation}
    \frac{|A(0,0,L,L/v_g)|^2}{|A(0,0,0,0)|^2} = \exp{(-(\delta k_z(\omega_c))^2 L^2)} = 1/2 \nonumber
\end{equation}

\noindent Therefore, the propagation distance \(L\) is,

\begin{equation}
    L = \frac{1}{\sqrt{2} \delta k_z(\omega_c)} \approx \frac{1}{\sqrt{2}}\frac{c}{\delta\omega} \frac{\cos{(\theta_c)}}{|1-c/v_g \cos{(\theta_c)}|} \label{eqns13}
\end{equation}

\noindent where \(\delta k_z(\omega_c) = \frac{\delta\omega}{c \cos{(\theta_c)}} \left(1- \frac{c}{v_g} \cos{(\theta_c)} \right)\), which is derived in the Supplementary Information Section \ref{S1}I. In the paraxial regime (\(\theta_c\ll1\)), the Eq. \ref{eqns13} becomes,

\begin{equation}
    L \approx \frac{1}{\sqrt{2}} \frac{c}{\delta\omega} \frac{1}{|1-c/v_g|}.\label{eqns14}
\end{equation}

\noindent The Eq. \ref{eqns14} can be compared to the expression \(L = c/\delta\omega|1-c/v_g|\) derived in the reference [S1] for STWPs in the paraxial regime. The factor of \(1/\sqrt{2}\) difference results from the different definitions of propagation distance in our manuscript and the reference. Ref. \cite{Yessenov_OE2019}%[S1]
defines the propagation distance in terms of a walk-off between the wavepacket and the pilot wave. Here we define the propagation distance in terms of the energy attenuation. 

\begin{figure}[b]
\centering
\includegraphics[width=0.25\linewidth]{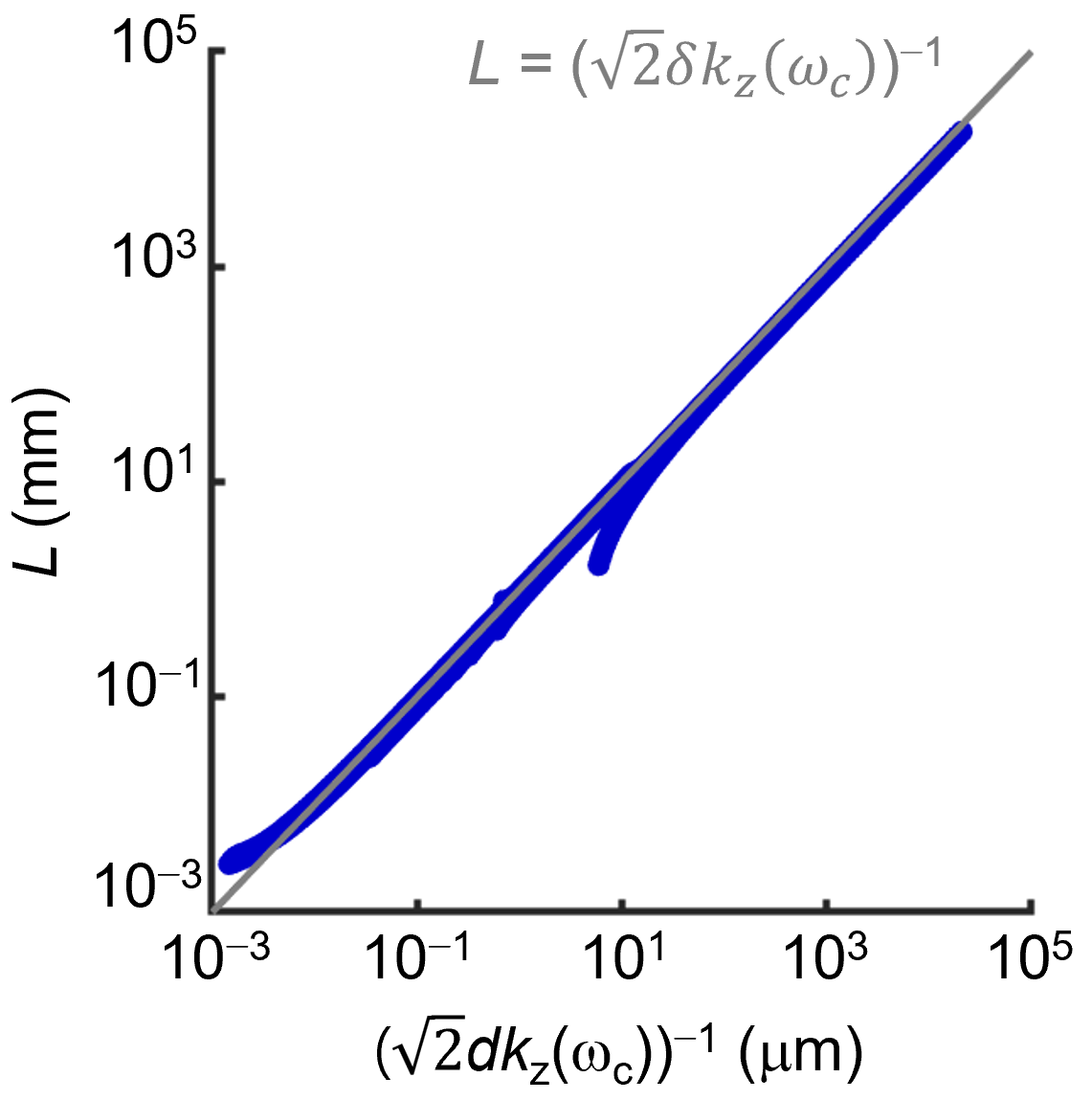}
\caption{Comparison between the propagation distance \(L\) in Figure 4b in main manuscript and analytic expression \((\sqrt{2}\delta k_z(\omega_c))^{-1}\). The gray line indicates the condition of \(L = (\sqrt{2}\delta k_z(\omega_c))^{-1}\).}
\label{figS2}
\end{figure}

As a validation of Eq. \ref{eqns13}, we plot the relation between \(L\) and \((\sqrt{2}\delta k_z(\omega_c))^{-1}\) in Figure \ref{figS2} as blue dots for each \(v_g^{(i)}/c\) and \(M\) used in Figure 4b in main manuscript. The numerical results agree excellently with Eq. \ref{eqns13} (We note that to apply in Eq. \ref{eqns13} for our case the \(v_g\) in Eq. \ref{eqns13} is the \(v_g^{(f)}\) for each wavepacket).

\noindent    
% \bibliography{Optica-journal-template}

% \noindent [S1] Yessenov, M., et. al., What is the maximum differential group delay achievable by a space-time wave packet in free space? Optics Express, 9(27), 12443-12457, 2019

\end{document}